\documentclass[aps,prl,twocolumn,superscriptaddress,english]{revtex4-1}

\usepackage[T1]{fontenc}
\usepackage{babel}
\usepackage{amsmath}
\usepackage{amssymb}
\usepackage{wasysym}
\usepackage{graphicx}
\usepackage{xcolor}
\usepackage{braket}
\usepackage{multirow}

\usepackage{hyperref}

\newcommand{\sh}{SH$_3$}
\newcommand{\ph}{PH$_3$}

\newcommand{\tc}{$T_\text{c}$}

\newcommand{\ep}{\textit{e-ph}~}

\begin{document}

\title{Superconductivity in sodalite-like yttrium hydride clathrates}

\author{Christoph Heil} 
%\affiliation{\Graz}
\affiliation{Institute of Theoretical and Computational Physics, Graz University of Technology, NAWI Graz, 8010 Graz, Austria}
\author{Simone di Cataldo}
\affiliation{Institute of Theoretical and Computational Physics, Graz University of Technology, NAWI Graz, 8010 Graz, Austria}
\author{Giovanni B. Bachelet}            
\affiliation{Dipartimento di Fisica, Sapienza Universit\`a di Roma, 00185 Roma, Italy}
\author{Lilia Boeri} \email{lilia.boeri@uniroma1.it}
\affiliation{Dipartimento di Fisica, Sapienza Universit\`a di Roma, 00185 Roma, Italy}

\date{\today}

\begin{abstract}
We report {\em ab-initio} calculations of the superconducting properties of two high-$T_\text{c}$ sodalite-like clathrate yttrium hydrides, YH$_6$ and YH$_{10}$, within the fully anisotropic ME theory, including Coulomb corrections. For both compounds we find almost isotropic superconducting gaps, resulting from a uniform distribution of the electron-phonon coupling over phonon modes and electronic states of mixed Y and H character. The Coulomb screening is rather weak, resulting in a Morel-Anderson pseu\-do\-po\-ten\-tial $\mu^* = 0.11$, at odds with claims of unusually large $\mu^*$ in lanthanum hydrides. The cor\-re\-spon\-ding critical temperatures at 300~GPa exceed room temperature ($T_\text{c} = 290$~K and 310~K for YH$_6$ and YH$_{10}$), in  agreement with a previous isotropic-gap calculation. The different response of these two compounds to external pressure, along with a comparison to low-$T_\text{c}$ superconducting YH$_3$, may inspire strategies to improve the superconducting properties of this class of hydrides.
\end{abstract}

\pacs{~}
\maketitle

The report of a superconducting  critical temperature ($T_\text{c}$) of 265~K in the lanthanum superhydride LaH$_{10}$ at 190~GPa~\cite{La_EXP1, La_EXP2, La_EXP3} set a new record for superconductivity only three years after another superhydride, \sh, 
%with a $T_\text{c}$ of 203~K, 
opened up the high-pressure route to conventional high-$T_\text{c}$ superconductivity~\cite{DrozdovEremets_Nature2015, Duan_SciRep2014}. These breakthroughs stem from two seminal papers of Neil Ashcroft, who first conjectured that high-$T_\text{c}$ conventional superconductivity would arise in high-pressure elemental metallic hydrogen~\cite{Ashcroft_1968}, and later proposed that the huge threshold pressure for hydrogen metalization might be significantly reduced in binary hydrogen compounds XH$_n$, by exploiting the additional internal pressure due to the X atoms~\cite{Ashcroft_2004}.

Three years of %theoretical and experimental
 research resulted in the determination of the high-pressure phase diagrams of most binary hydrides~\cite{Oganov_review,Zurek_review}, clarifying that those hydrides exhibiting high-$T_\text{c}$ superconductivity mainly fall into two classes: ($i$) covalent hydrides, like \sh \ and \ph, in which H and the other element X form a network of covalent bonds, driven metallic by the high pressure, and ($ii$) metallic hydrides of alkaline and rare earths, like LaH$_{10}$, which form  hydrogen-rich sodalite-like clathrates (SLC) with highly symmetric structures~\cite{La_EXP1, La_EXP2, La_EXP3, Hclathrates_Wang_PNAS2012, Yhydrides_Li_Scirep_2015, Liu_PNAS_la_hydrides, MA_REHX_PRL_2017}, whose $T_\text{c}$'s are close to, or even higher than room temperature. In class  ($i$), the chance of high-$T_\text{c}$ superconductivity is governed by the degree of covalency of the H--X bonds, and X=S seems to approach a {\em sweet spot}~\cite{Duan_SciRep2014, SH_PRB-Mazin-2015, Heil-Boeri_PRB2015, Flores-Livas2016, PhysRevLett.114.157004, Drozdov_PH3_arxiv2015, Flores_PH3_PRBR2016, shamp_decomposition_2015, Fu_Ma_pnictogenH_2016};  in class  ($ii$), the specific electron-phonon mechanism leading to high-$T_\text{c}$ has not yet been identified as clearly~\cite{La_TH_oganov,La_TH_topological,LA_TH_Naumov_quantumlattice}.

Since in the compounds with  highest $T_\text{c}$ the H-H distance is close to that of solid hydrogen \cite{PhysRevB.84.144515,PhysRevB.93.174308}, many authors emphasize the role of the H sub\-lattice and regard LaH$_{10}$ as the first experimental evidence of high-$T_\text{c}$ superconductivity in precompressed atomic hydrogen. According to this picture, once the X atoms provide charge to the hydrogen sublattice and sintuitabilize a crystal structure with sufficiently small H-H distances, high-$T_\text{c}$ superconductivity follows. Recently, however, such an oversimplification lead to wrong expectations \cite{Pepin_FEH_2017,Heil_FEH}.
In fact, the pre-requisite for high-$T_\text{c}$  in high-pressure hydrides
is a substantial role of H %electronic and vibrational 
states in superconductivity, which cannot be guessed based on
 H-H distances alone. Our aim is to identify the  electronic structure features 
behind high-$T_\text{c}$ superconductivity in high-pressure SLC hydrides.

We will re-examine two representative \hbox{high-$T_{\rm c}$,} high-pressure hydrides of this class, YH$_6$ and YH$_{10}$, using the fully anisotropic {\em ab-initio} Migdal-Eliashberg theory as implemented in the \textsc{Epw} code~\cite{foot_CD,margine_anisotropic_2013,ponce_epw:_2016}, with the aim of identifying a rationale on the physico\-chemical ingredients needed to reduce their stabilization pressure without reducing their high $T_\text{c}$. In this light, we do not address the thermo\-dynamics of the Y-H system, already  analyzed by previous works, and concentrate on the high-symmetry SLC structures of YH$_6$ and YH$_{10}$ which, according to {\em ab-initio} calculations, are stable with record $T_\text{c}$'s of 260~K for YH$_6$ at 120~GPa, and of 303~K for YH$_{10}$ at 400 GPa~\cite{MA_REHX_PRL_2017,Liu_PNAS_la_hydrides,foot_YH9}.
Since in the Periodic Table yttrium belongs to the same group as lanthanum (one row above),  the crystal structures and  super\-conducting properties of its high-pressure  hydrides closely track the parallel lanthanum compounds~\cite{La_EXP1,La_EXP2,La_EXP3}. The practical advantage of yttrium is that its $f$~states, way above the Fermi level, play no role in the bonds and bands of its hydrogen compounds. In lanthanum compounds, instead, the $f$ states (troublesome both for density-functional  and pseudopotential theory) are near the Fermi level and must be included, although in the end their contribution to stability and superconductivity turns out to be negligible~\cite{La_TH_topological}.

\begin{figure}[t]
	\includegraphics[width=0.80\columnwidth]{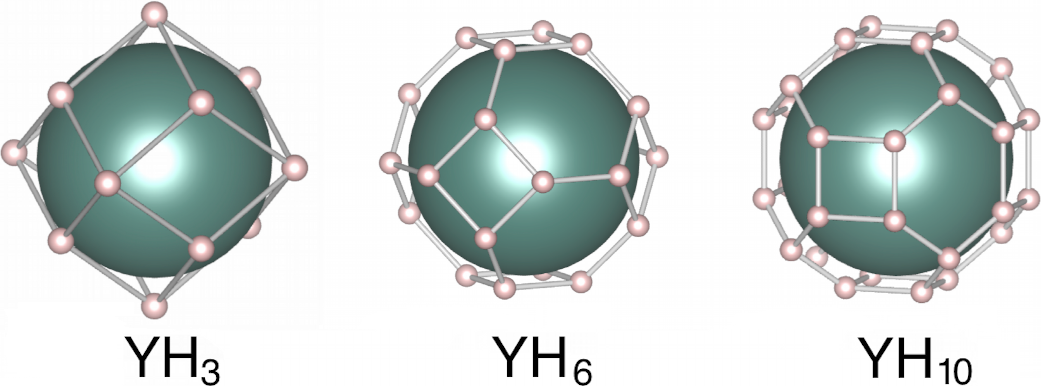}
	\caption{
	The crystal structures of fcc YH$_{3}$ (left), bcc YH$_{6}$ (center), and fcc YH$_{10}$ (right) appear as space-filling polyhedral hydrogen cages (H$ = $small pink balls) with an yttrium atom (Y$ = $large green balls) in their middle, whose radius, for visual clarity, was chosen equal to 1.62~\AA (midway between core and covalent radius). In this picture the \hbox{H-H} distan\-ces corre\-spond to an external pressure of 300~GPa: 
 	$d_{\rm HH} = 1.74$~\AA \ for YH$_{3}$, $d_{\rm HH} = 1.19$~\AA \ for YH$_{6}$, and two slightly different lengths $d_{\rm HH} = 1.03$, 1.11~\AA \ for YH$_{10}$ (see text).
 	} 
	\label{fig:struc}
\end{figure}

Our results confirm that, in addition to a reasonably small H-H distance, both the superconducting behavior and the dynamical stability under pressure  of these YH$_n$ hydrides are determined by the peculiar geometry of such a densely connected H lattice (similar to a {\em sponge} of H filaments whose cavities are occupied by Y atoms), and not by the chemical details of the enclosed atom.

Fig.~\ref{fig:struc} shows the two high-$T_{\rm c}$ yttrium hydrides considered in this work~\cite{foot_SM_crystal}, YH$_6$ and YH$_{10}$, together with the  low-$T_{\rm c}$ YH$_3$ crystal,  experimentally observed above 10 GPa, whose predicted maximum $T_{\rm c}$ is 40~K at 18~GPa \cite{YH3_exp_PRB2007,YH3_theory_PRL2009}. In YH$_3$ and YH$_6$ a hydrogen atom sits on each of the 14 (24) vertices of the fcc (bcc) Wigner-Seitz primitive cell; in YH$_{10}$ it sits on each of the 32 vertices of a chamfered cube. For each such polyhedron well-known relations connect the edge length (the H-H distance $d_{\rm HH}$), the volume $V$ (the unit cell volume of the corresponding crystal), the average radius, etc. For example $d_{\rm HH}\!=\!0.69 \, V^{1/3}$ in YH$_3$, $0.45 \, V^{1/3}$ in YH$_6$, and $0.38 \, V^{1/3}$ in YH$_{10}$. Geometrical constraints not only control ($i$) the H-H distance, important for high-$T_{\rm c}$, but also ($ii$) the Y-H distance, important for the involvement of Y in the \ep interaction, and ($iii$) how tight or loose is the host clathrate cavity where the (fixed-size) guest atom sits; which, in turn, triggers the onset of their dynamical instability at ``low'' pressure, discussed later ~\cite{foot_SM_crystal}.

\begin{figure}[t]
	\includegraphics[width=0.90\columnwidth]{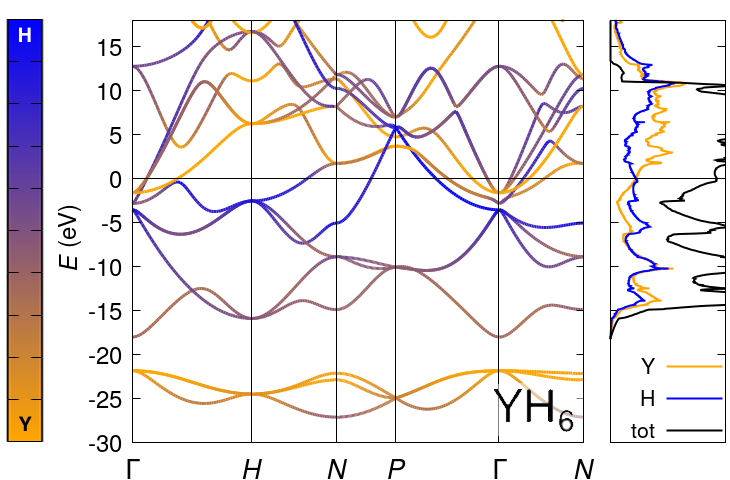}\\
	\includegraphics[width=0.90\columnwidth]{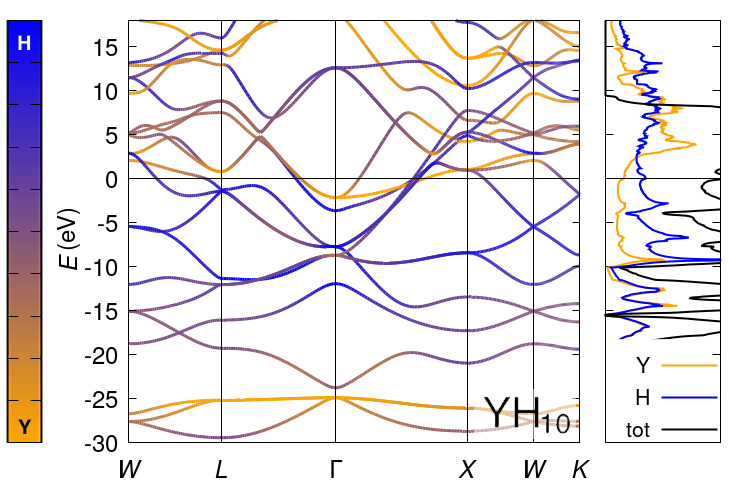}
	\caption{
	Left: electronic energy bands, where the color gradient indicates the projection onto H (blue) and Y (orange) states.
	Right: total DOS (black), partial Y DOS (orange), and partial H DOS (blue). Energies are referred to E$_F$.
	}
	\label{fig:bands}
\end{figure}

We now focus on the two high-$T_{\text{c}}$ superconductors~\cite{foot_YH3}, whosebands (left) and densities of states (DOS, right) are shown in Fig.~\ref{fig:bands} for YH$_6$ (top) and YH$_{10}$ (bottom).
%we will come back to the \hbox{low-$T_{\text{c}}$} YH$_3$  later on.

\begin{figure}[!hb]
    \includegraphics[height=0.232\textheight]{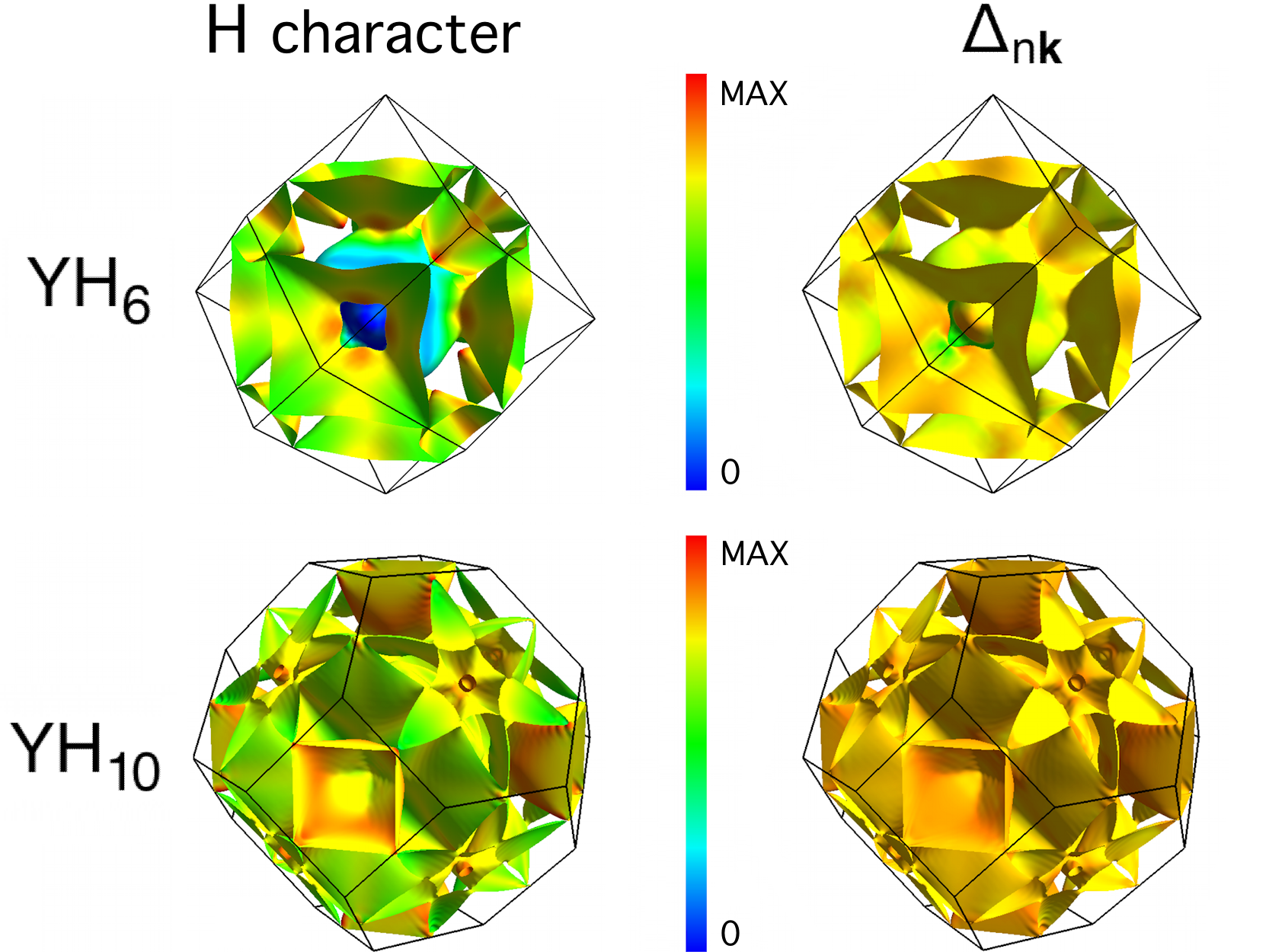}
    \caption{
    Fermi surfaces of YH$_{6}$ (top row) and YH$_{10}$ (bottom row). In the left panels the color scale spans the projection onto H states, where blue corresponds to 0 and red to 1; in the right panels it spans the values of the anisotropic gap function at 40~K, blue being 0 meV and red the maximum  of 59 (74)~meV for YH$_6$ and YH$_{10}$, respectively. The \hbox{band-by}-band decomposition (individual Fermi-surface sheets) is available in Sec.~II of the Supplemental Material, Tab.~S3 and S4~\cite{SM_link}.
}
    \label{fig:FS}
\end{figure}

Unless otherwise stated, all subsequent results refer to a pressure of 300~GPa, where both YH$_6$ and YH$_{10}$ are dynamically stable, and their $T_\text{c}$ is close to its maximum. The color gradient indicates the projection onto H (blue) and Y (orange) states. In both compounds the hydrogen-derived bands have a total bandwidth of $\sim 40$~eV. Remarkably, by taking into account the materials' lattice geometries, their dispersion over this energy range is well described by quasi-free-electron bands~\cite{Simone_paper}, with largest deviations where the H- and Y-derived states significantly hybridize, i.e., $\sim 25$~eV below the Fermi level ($4p$ semicore states) and in a region of $\sim 10$~eV around the Fermi level ($4d, 5s$ states).
The Fermi level cuts the band structure where both H and Y contributions to the electronic structure are sizable: In particular, around the Brillouin zone center ($\Gamma$) the bands have mostly Y character, while at its boundaries they are mostly H~\cite{foot_SM_fermi}. The corresponding Fermi surfaces are shown in Fig.~\ref{fig:FS}.\par
\par In a superconductor, when two or more orbitals/bands at the Fermi surface couple to phonons with \hbox{different} intraband strengths, an an\-iso\-tro\-pic superconducting gap $\Delta_\mathbf{n k}$ results. Its behavior can be obtained entirely from first principles within the anisotropic Migdal-Eliashberg (ME) theory: The anisotropic \ep Eliash\-berg functions are calculated within the linear-response theory, using the Wannier interpolation technique imple\-men\-ted in the \textsc{Epw} code \cite{margine_anisotropic_2013,ponce_epw:_2016} and the $GW$ approximation for the fully screened Coulomb interaction~\cite{giustino_gw_2010, lambert_ab_2013}. We do this for the first time for YH$_{6}$ and YH$_{10}$, showing our result in the right panels of Fig.~\ref{fig:FS}. 

Before we comment this figure, let us discuss the main features of the phonon spectra and Coulomb interaction~\cite{migdal_interaction_1958, eliashberg_interactions_1960, foot_SM_decomposition}. In both YH$_{6}$ and YH$_{10}$ the Eliash\-berg spectral function (Fig.~S3-S4 in the Supplemental Material~\cite{SM_link}) shows a rather uniform distribution of the \ep coupling over all phonons, including the low-energy modes which are essentially of Y character. Compared to YH$_6$, the shorter, stiffer H-H bonds of YH$_{10}$ translate into 20\% larger frequencies for the high-energy, bond-stretching modes. The average \ep matrix elements are also higher, leading to a larger \ep coupling  in YH$_{10}$ ($\lambda\!=\!2.41$) than in YH$_6$ ($\lambda\!=\!1.73$). According to our calculations, the Coulomb pseudopotential is the same in both compounds: $\mu^*\!=\!0.11$, resulting from a $GW$-screened Coulomb interaction $\mu_c\!=\!0.11$ and a negligible Morel-Anderson renormalization. This is, to our knowledge, the first {\em ab-initio} estimate of Coulomb screening in H clathrates; the value $\mu^*\!=\!0.11$ places these compounds in the same ballpark as most conventional metals. It is reasonable to assume that similar values of $\mu^*$ occur  in SLC hydrides formed by other metals as well. On this basis the anomalously large $\mu^*\simeq 0.22$ invoked in Ref.~\onlinecite{La_TH_oganov} to theoretically reproduce the experimental $T_\text{c}$ appears unlikely.

Back to Fig.~\ref{fig:FS}, we observe that while in YH$_6$ and YH$_{10}$ the distribution of Y and H character on the Fermi surface is uneven (left panels), this only yields minor ($\pm 10 \%$) fluctuations of the superconducting gap around its average value (right). This quasi-isotropic gap is restored by the strong Y-H interorbital interactions due to the compact, quasi-spherical geometry of the system: All lattice vibrations, including bending and breathing modes of the cages, modulate the Y-H distance and thus the overlap between Y and H orbitals, which, in turn, washes out most anisotropic effects on superconductivity \cite{BCS_2bands}.
\begin{figure}[t]
    \includegraphics[width=1\columnwidth]{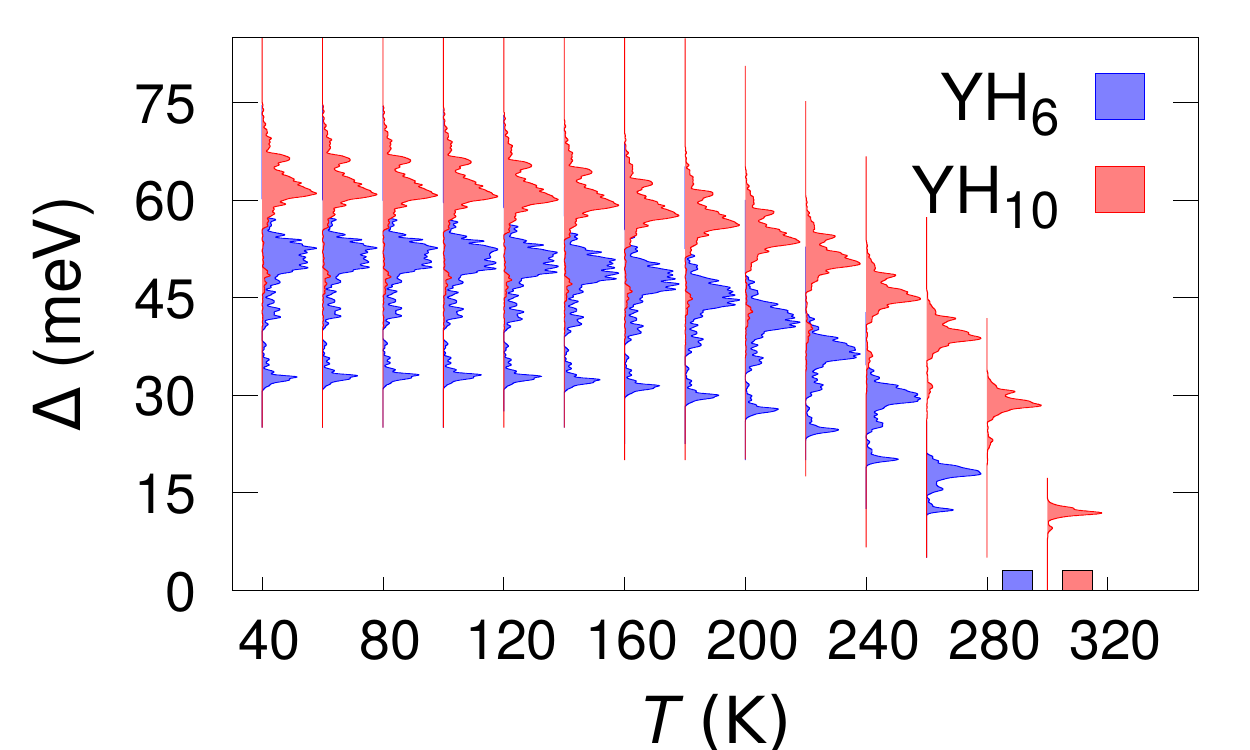}
    \caption{
     Energy distribution of the superconducting gap for YH$_{6}$ (blue) and YH$_{10}$ (red) as a function of temperature. The rectangles show the extrapolated $T_\text{c}$ values. 
    }
    \label{fig:delta_vs_T}
\end{figure}

\begin{figure}[t]
	\includegraphics*[width=0.9\columnwidth]{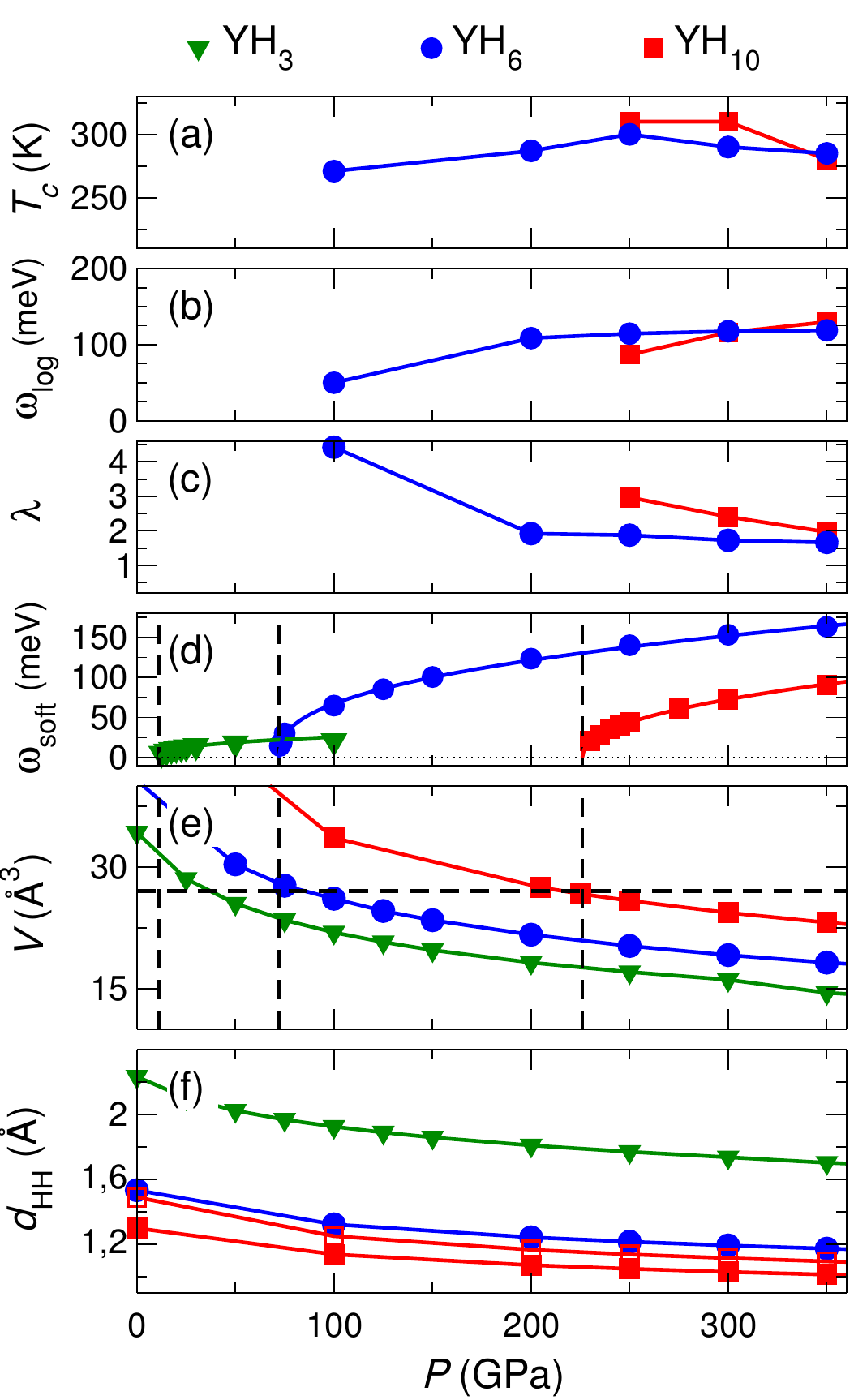}
	\caption{
		Behavior of several properties of YH$_3$ (green triangles), YH$_{6}$ (blue circles) and YH$_{10}$ (red squares) as a function of pressure. 
		($a$) \tc \ from anisotropic ME equations ($\mu^*=0.11$);
		($b$)-($c$) Momenta of the \ep spectral function $\alpha^2 F(\omega)$, $\omega_{\log}$ and $\lambda$.
		($d$) Frequency of the soft mode.
		($e$) Volume of the Wigner-Seitz unit cell.
        ($f$) Nearest-neighbor H-H distance ($d_{\rm HH}$), see Fig.~\ref{fig:struc}. The dashed lines in ($d$) and ($e$) indicate the points where the SLC structures become dynamically unstable.
	}
	\label{fig:pressure}
\end{figure}

We studied the temperature dependence of the superconducting gap by solving the an\-iso\-tro\-pic ME equations at different temperatures; Fig.~\ref{fig:delta_vs_T} displays the temperature evolution of its energy distribution function over the Fermi surface. Well below  $T_{\rm c}$, i.e. for $T\!<\!80$\,K in Fig.\,\ref{fig:delta_vs_T}, this distribution is nearly indepen\-dent of temperature, and shows a broad maximum around 65~meV (55~meV) for YH$_{10}$ (YH$_6$), originating from the two zone-boundary Fermi surfaces and the two large zone-center Fermi surfaces, plus a smaller tail at lower energies (52~meV for YH$_{10}$ and 36~meV for YH$_6$), due to the two smallest zone-center Fermi surfaces (see Fig.~\ref{fig:FS} and Supplemental Material~\cite{SM_link}). The gap closes at a critical temperature of 290 K in YH$_6$ and 310 K in YH$_{10}$~\cite{foot_AD}. Since in both compounds the dependen\-ce of $T_{\rm c}$ on pressure is very weak, as shown in panel (a) of Fig.~\ref{fig:pressure}, our predictions for $T_\text{c}$ amount to a remarkable agreement with Ref.~\cite{MA_REHX_PRL_2017}, which, using the isotropic Migdal-Eliashberg theory and $\mu^*$= 0.10, estimated 264~K for YH$_6$ at 120~GPa and 303~K for YH$_{10}$ at 400~GPa.

As shown in panels (b) and (c) of Fig.\,\ref{fig:pressure}, the weak pressure de\-pen\-den\-ce of $T_{\rm c}$ results from an almost perfect compensation between the average phonon energy $\omega_{\log}$, which increases with pressure~\cite{foot_SM_phonons}, and the \ep coupling constant $\lambda$, which, instead, decreases. For both compounds this balance approximately holds down to  a pressure of $\sim$250~GPa, below which the lowest optical branch ($\Gamma\!-\!L$ line in YH$_{10}$,  $\Gamma\!-\!H$ line in YH$_6$) gets softer and softer, eventually leading to a dynamical instability at $\sim$226 GPa and $\sim$72 GPa, respectively \cite{foot_SM_phonons}. The soft branch carries a substantial fraction of the total \hbox{\ep} coupling, but a glance at the $\mathbf{q}$-dependent electronic susceptibility~\cite{heil_accurate_2014, foot_SM_phonons} shows that its softening is not due to nesting and must be related to the \ep matrix elements. 

This, in turn, suggests an intrinsic instability of  the Y-H system in the SLC structure, which is robust against minor changes of the electronic structure. The common physical origin of the instability of YH$_{10}$ and YH$_6$ at two very different critical pressures is revealed by their comparison with yet another SLC yttrium hydride: YH$_3$ (green triangles in Fig.\,\ref{fig:struc}), which, according to our calculations, remains stable down to the much lower pressure of 11.5 GPa \cite{foot_YH3}.
Panel~(e) of Fig.~\ref{fig:pressure} shows the $V$ vs $P$ equation of state for the three compounds, and clearly evince that the three different pressures below which the soft modes become imaginary in YH$_3$ (green), YH$_6$ (blue), YH$_{10}$ (red),  correspond to a single volume  of $\sim 27$~\AA$^3$, which, in fact, equals the volume of a sphere of radius $\sim 1.9$~\AA, the covalent radius of Y.

This suggests that, for SLC hydrides with chemical formula XH$_n$, the minimum stabilization pressure is dictated by the size of the guest atom X: When the size of the primitive cell exceeds it, the hydrogen cage becomes too loose to constrain this atom in its middle, and hence the H lattice breaks down. If this is true, then, for a given atom X, the compounds with larger $n$ (implying denser hydrogen cages with smaller H-H distances) will require larger stabilization pressures. So, as far as the {\em dynamical stability} is concerned, cages with small $n$ and large H-H distances $d_{\rm HH}$ are preferable, because they require lower pressures; on the other hand, 
%as already recognized by several authors and confirmed by this work, 
besides a substantial contribution of H electronic and vibrational states to superconductivity, the high-\tc \ hydrogen  superconductivity needs small H-H distances (close to the shortest atomic-solid-hydrogen value $d_{\rm HH}=0.98 \AA$ at 500 GPa)~\cite{MA_REHX_PRL_2017,Liu_PNAS_la_hydrides}, and thus large $n$.

In other words, the competing requirements for dynamical stability and superconductivity, together with the different geometrical prefactor which affect the dependence of $d_{\rm HH}$ on the primitive cell/cage volume $V$  \cite{foot_SM_crystal}, provide a natural explanation, pictorially summarized by panels (e) and (f) of Fig.~\ref{fig:pressure}, why YH$_6$ (intermediate cage volume, small $d_{\rm HH}$) is better than both YH$_3$ (smallest cage volume, but too large $d_{\rm HH}$, almost twice than in atomic-solid-hydrogen up to 350 GPa)  and YH$_{10}$ (smallest $d_{\rm HH}$, but too large a cage volume).

In summary, we have studied the superconducting properties of two high-pressure yttrium hydrides, YH$_6$ and YH$_{10}$, using first-principles anisotropic Migdal-Eliashberg theory, including Coulomb corrections. Our calculations confirm the room-temperature superconductivity found by other authors, and show that it results from a strong \ep interaction which is rather uniformly spread over electronic and vibrational states of both hydrogen and yttrium sublattices. The Coulomb pseudopotential parameter, which for these compounds we computed for the first time \emph{ab-initio} within the $GW$ approximation, is in line with the values found in most conventional superconductors ($\mu^*$= 0.11), in contrast to recent studies, which propose a much larger value for related lanthanum SLC hydrides~\cite{La_TH_oganov}.
%Our results show that 
Due to the peculiar geometry by which the yttrium SLC hydrides implement a dense hydrogen lattice, optimizing their superconducting behavior under pressure
requires a careful compromise between H packing and structural stability. Our findings may inspire optimization strategies for other superconducting hydrides
of the same class.

\begin{acknowledgments}
 This work was supported by the Austrian Science Fund (FWF) Projects No. J 3806-N36 and P 30269-N36,  the dCluster of the Graz University of Technology and the VSC3 of the Vienna University of Technology. L.B. and G.B.B. acknowledge support from Fondo Ateneo-Sapienza 2017.
 L.B. would like to thank Antonio Sanna for useful discussions on Migdal-Eliashberg theory.
\end{acknowledgments}

%\bibliographystyle{apsrev4-1}
%\bibliography{paper}%merlin.mbs apsrev4-1.bst 2010-07-25 4.21a (PWD, AO, DPC) hacked

\end{document}